\documentclass[titlepage]{csetr}

\usepackage{graphicx}

\usepackage{setspace}
\usepackage{geometry}
\usepackage{float}
\usepackage{setspace,supertabular}
\usepackage{longtable}
\usepackage[usenames]{color}
\usepackage{verbatim}
\usepackage{subfigure}
\usepackage{amsmath}
\usepackage[table]{xcolor}

\renewcommand{\and}{\hspace{.5cm}}

\trnumber{1116}
\trdate{September 2011}

\title{%
  Temporal Provenance Model (TPM):\\Model and Query Language
}
\author{%
  Seyed-Mehdi-Reza Beheshti$^1$ \and %
  Hamid Reza Motahari-Nezhad$^2$ \\ %
  Boualem Benatallah$^1$\\[2em]
  $^1\, $University of New South Wales\\ Sydney 2052, Australia \\%
  \email{\{sbeheshti,boualem\}@cse.unsw.edu.au}\\ \\
  $^2\,$HP Labs Palo Alto\\ CA 94304, USA \\%
  \email{hamid.motahari@hp.com}\\[3cm]
}

\date{}
\begin{document}
\maketitle

\begin{abstract}
Provenance refers to the documentation of an object's lifecycle. This documentation (often represented as a graph) should include all the information necessary to reproduce a certain piece of data or the process that led to it. In a dynamic world, as data changes, it is important to be able to get a piece of data as it was, and its provenance graph, at a certain point in time. Supporting time-aware provenance querying is challenging and requires: (i)~explicitly representing the time information in the provenance graphs, and (ii)~providing abstractions and efficient mechanisms for time-aware querying of provenance graphs over an ever growing volume of data. The existing provenance models treat time as a second class citizen (i.e. as an optional annotation). This makes time-aware querying of provenance data inefficient and sometimes inaccessible. We introduce an extended provenance graph model to explicitly represent time as an additional dimension of provenance data. We also provide a query language, novel abstractions and efficient mechanisms to query and analyze timed provenance graphs. The main contributions of the paper include: (i)~proposing a Temporal Provenance Model (TPM) as a timed provenance model; and (ii)~introducing two concepts of \emph{timed folder}, as a container of related set of objects and their provenance relationship over time, and \emph{timed paths}, to represent the evolution of objects tracing information over time, for analyzing and querying TPM graphs. We have implemented the approach on top of FPSPARQL, a query engine for large graphs, and have evaluated for querying TPM models. The evaluation shows the viability and efficiency of our approach.
\end{abstract}

\section{Introduction}

Provenance refers to the documented history of an object (e.g. documents, data, and resources) or the documentation of processes in an object's lifecycle, which tracks the steps by which the object was derived~\cite{provSurvey3} and evolved. This documentation (often represented as graphs) should include all the information necessary to reproduce a certain piece of data or the process that led to that data~\cite{Moreau:OPM}. The ability to analyze provenance graphs is important as it offers the means to verify data products, to infer their quality, and to decide whether they can be trusted~\cite{ProvenanceReproducibility}. In a dynamic world, as data changes, it is important to be able to get a piece of data as it was, and its provenance graph, at a certain point in time. Under this perspective, the provenance queries may provide different results for queries looking at different points in time. Enabling time-aware querying of provenance information is challenging and requires: (i)~explicitly representing the time information in the provenance graphs, and (ii)~providing abstractions and efficient mechanisms for time-aware querying of provenance graphs over an ever increasing volume of~data.

The existing provenance models, e.g., the open provenance model (OPM)~\cite{OPM2010,Moreau:OPM}, treat time as a second class citizen (i.e. as an optional annotation of the data) which will result in loosing semantics of time and makes querying and analyzing provenance data for a particular point in time inefficient and sometimes inaccessible. For example, annotations assigned to an entity (e.g. a file or Web resource) today may no longer be relevant to the future representation of that entity, as entity attributes are very likely to have different states over time and the temporal annotations may or may not apply to these evolving states. Due to the implicit treatment of time, abovementioned approaches do not enable explicit representation of the evolution of relevant subgraphs (i.e. \emph{group} of interrelated objects) and paths (i.e. discovering historical paths through provenance graphs forms the basis of many provenance queries~\cite{RDFProv,provenanceQuery1,provQL}) over time. For example, the shortest path between an object to its origin may change over time~\cite{graphEvolveVLDB} as provenance metadata forms a large, dynamic, and time-evolving graph.

To address these challenges, there is a need for explicit representation of temporal information in provenance models, and also efficient approaches for analyzing and querying provenance information with respect to time. In this paper, we introduce a graph data model for provenance called Temporal Provenance Model (TPM), and a query language to query and analyze temporal provenance graphs. We introduce two concepts of \emph{timed folders}, i.e. a placeholder for a group of inter-related time-evolving entities (e.g. group of related activities such as process instances in the context of a business process~\cite{BPM11}), and \emph{timed paths}, i.e. a placeholder for the set of entities that are related to each other through transitive relationships. These relationships may evolve over time, e.g., to show the evolution of the historical path of an artifact to its origin. Timed folders can be used to: (i)~store and represent the result of provenance queries over TPM graphs, e.g.~retrieving the provenance graph of snapshots of objects or group of related objects over time; and (ii)~partition TPM graphs into various temporal phases which can simplify the discovery of temporal relationship among provenance graph entities over time. Timed paths can be used to store and represent the results of path queries over TPM graphs, e.g. derivation queries which retrieve historical path(s) from an object to its origin over time. Timed folders and paths show their evolution for the time period that they represent and can be used for further querying.

In summary, we present a novel framework for modeling, querying, and analyzing temporal provenance metadata. The unique contributions of the paper are as follows:

\begin{itemize}
  \item We propose a temporal provenance model (TPM), i.e. a temporal model for collecting and exchanging provenance metadata.

  \item We introduce two concepts of timed folders and timed paths, which help in analyzing provenance graphs. Folders enable grouping related entities and paths help in analyzing the history of entities in time. Timed folder and path nodes can show their evolution for the time period that they represent.

  \item We present a graph query language to query and analyze TPM graphs.

  \item We provide a front-end tool for assisting users to create provenance queries in an easy way and visualizing TPM graphs and query results.
\end{itemize}

The remainder of this paper is organized as follows: We present background and the related work in section~\ref{BackgroundRW}. Section~\ref{bgScenario} presents an example scenario. In section~\ref{SecMetaModel} we present the temporal provenance model (TPM). In section~\ref{SecQueryLanguage} we propose a query language for querying TPM graph. In section~\ref{Experiments} we describe the query engine implementation and evaluation experiments. Finally, we conclude the paper with a prospect on future work in Section~\ref{Conclusion}.

\section{Background and Related Work}
\label{BackgroundRW}

Prior work on modeling and representing provenance metadata~\cite{provSurvey1,provSurvey2,provSurvey3} (e.g. lineage, where-provenance, why-provenance, dependency-provenance, how-provenance, and provenance-traces models)  model provenance as a directed acyclic graph, where the focus is on modeling the process that led to a piece of data. They present vocabularies to model process activities and their causal dependency, i.e. the relationship between an activity (the cause) and a second activity (the effect) where the second activity is understood as a consequence of the first. For example, open provenance model (OPM)~\cite{Moreau:OPM, OPM2010} (which proposed to design a standard graph data model and vocabulary for provenance), presents graph nodes as \emph{data artifacts} (i.e.~defined as product of human intelligence and effort, and is classified as tangible,e.g. a physical object, and intangible, e.g. digital representation of an object), \emph{processes} (i.e.~action or series of actions performed on or caused by artifacts, and resulting in new artifacts~\cite{OPM2010}), and \emph{agents} (i.e.~contextual entity acting as a catalyst of a process, enabling, facilitating, controlling, affecting its execution~\cite{Moreau:OPM}, e.g. people and services). Also, five causal relationships are recognized in OPM: a process `used' an artifact, an artifact `was generated by' a process, a process `was triggered by' a process, an artifact `was derived from' an artifact, and a process `was controlled by' an agent.

In a dynamic world data changes, so the graphs representing data provenance evolve over time. It is important to be able to reproduce a piece of data or the process that led to that data for a specific point in time. This requires modeling \emph{time} as a first class citizen in the provenance models. Times, intervals, and versioning can be very important in understanding provenance graphs as the structure of such graphs evolves over time. Today's approaches in modeling provenance, e.g. in OPM, treat time as a second class citizen. Considering time as a first class citizen, will enable retrieving multiple snapshots of entities (versions) in the past which can help in capturing the provenance for each version of an entity independently. Moreover, it can help in understanding the role of each entity in the temporal context of the entire system. Today's approaches in modeling provenance, e.g. in OPM, treat time as a second class citizen and do not provide abstractions to consider group of related entities as first class objects (which can help in analyzing provenance graphs over ever increasing volumes of data). Pursuing these approaches, it will be cumbersome and lead to an increased query complexity to analyze and understand relationships among objects over time.

We study the related work into three main areas: provenance representation and querying, temporal graphs and databases, and graph query languages:

\subsection{Provenance Representation and Querying}

Many provenance models~\cite{provSurvey1,provSurvey2,provSurvey3,Moreau:OPM,provenanceModel2,provenanceModel1} have been presented in a number of domains (e.g. databases, scientific workflows and the Semantic Web), motivated by notions such as influence, dependence, and causality. Why-provenance \cite{provSurvey2,provSurvey1} models the influences that a source data had on the existence of the data. Where-provenance~\cite{provSurvey3} focuses on the dependency to the location(s) in the source data from which the data was extracted. How-provenance~\cite{provSurvey3,provSurvey2} represents the causality of an action or series of actions performed on or caused by source data. Open Provenance Model~\cite{Moreau:OPM} focuses on the causality dependency among data artifacts, the process which generates them, and contextual entity or entities acting as a catalyst of the process. PAPEL~\cite{provenanceModel2}, a provenance-aware policy definition model, extends OPM in order to specify the relation between policy condition and provenance information. Cao et. al.~\cite{provenanceModel1} proposed a two-layer model for representing provenance information capable of representing both execution information and higher level process details. These approaches consider provenance as an annotated causality graph. Using provenance models in these approaches leads to an increased query complexity in analyzing the relationships among data objects and their provenance metadata over time.

Discovering historical paths through provenance graphs forms the basis of many provenance query languagesqueries~\cite{provQL,provenanceQuery1,provenanceQuery2,RDFProv,EDBTprovQuery}. In ProQL~\cite{provQL} a query takes a provenance graph as an input, matches parts of the input graph according to path expression and returns a set of paths as the result of the query. PQL~\cite{provenanceQuery1} uses a semi-structured data model for handling provenance and extends Lorel query language for traversing and querying provenance graph. NetTrails~\cite{provenanceQuery2} proposes a declarative platform for interactively querying network provenance in a distributed system in which query execution performs a traversal of the provenance graph. RDFProv~\cite{RDFProv} is an optimized framework for scientific workflow provenance querying and management. Missie et. al.~\cite{EDBTprovQuery2} present a provenance model and query language for collection-oriented workflow systems (e.g. Taverna). They emphasize on querying the provenance of collection of activities. These related activities are not considered as first class objects in the proposed graph. Moreover, they do not support modeling, querying and analyzing the evolution of group of related entities over~time.

\subsection{Temporal Databases and Graphs}

A recent book~\cite{TemporalDatabase} discusses the time domain, and the termporal data representation and mining approaches. It argues that time is an important aspect of all real-world data and it is critical to model temporal dimension of data. Considering time as an additional dimension in data will directly affect the process that led to that data, i.e. its provenance. Temporal databases enable retrieving multiple snapshots (versions) of data artifacts at different points in time. However, a temporal database does not capture important information for data provenance such as activities performed on the data, agents acting on the data, and the relationships that the different versions of artifacts have to each other in various points in time. Annotation techniques~\cite{annotationGraph,timestampGraph,graphAnnotation2} represent another perspective to model temporal relationships. In this technique, system entities are (optionally) labeled with time offsets. These systems consider well-designed systems of interacting components and analyze data within a specific narrow domain. Using temporal annotations will result in loosing the semantics of time, as the timestamp of a data would capture the state and time for a snapshot but not the temporal evolution history of the artifact along with the version of the artifact at each point in time~\cite{TemporalNetwork,TemporalRDF}.

In recent years, a plethora of work~\cite{TemporalNetwork,Kostakos:TemporalGraph,graphEvolveVLDB} has focused on temporal graphs since in many applications, information is best represented and stored as graphs. The focus of temporal graphs is to model evolving, time-varying, and dynamic networks of data. They capture a snapshot for various states of the graph over time. For example, Ren et. al.~\cite{graphEvolveVLDB} propose a historical graph-structure to maintain analytical processing on such evolving graphs. Moreover, authors in~\cite{Kostakos:TemporalGraph,graphEvolveVLDB} propose approaches to transform an existing graph into a similar temporal graph to discover and describe the relationship between the internal object states. In our approach, we propose a temporal provenance model to capture the provenance of time-sensitive data where this data can be modeled as temporal graph. We also provide abstractions and efficient mechanisms for time-aware querying of temporal provenance graphs.

\subsection{Graph Query Languages}
\label{GraphQueryLanguages}

Approaches for modeling and querying graphs (e.g.~\cite{EPSPARQL,T_SPARQL,SPARQLST,TemporalRDF}) provide temporal extensions of existing graph models and languages. Tappolet et. al.~\cite{TemporalRDF} provide temporal semantics for RDF graphs, i.e. a finite set of RDF triples (subject, predicate, object) where RDF (a W3C standard) is a framework for representing information in the Web. The authors follow the concepts of temporal databases to introduce time as a new dimension in RDF graphs. For querying temporal graphs, they propose $\tau$-SPARQL, i.e. an extension of SPARQL~\cite{SPARQL} which is a declarative query language (a W3C standard) for extracting information from RDF graphs. Grandi~\cite{T_SPARQL} presents another temporal extension for SPARQL, i.e. T-SPARQL, aimed at embedding several features of TSQL2~\cite{TemporalDatabase} (temporal extension of SQL). SPARQL-ST~\cite{SPARQLST} is an extension of SPARQL for complex spatial-temporal queries which associates time intervals with RDF statements to realize temporal RDF graphs. EP-SPARQL~\cite{EPSPARQL} is an extension of the SPARQL supporting real time detection of temporal complex patterns in stream reasoning. Our work differs from these approaches as we enable registering a time-sensitive query once, propose timed abstractions (i.e. folders and paths) to store the result of such queries, and enable analyzing the evolution of such timed abstractions over time.

In FPSPARQL~\cite{BPM11}, our previous work, we extend SPARQL to support querying grouped entities and discovering paths in graph. In this work, we extend FPSPARQL to support temporal queries (e.g. querying time, intervals, and versions), and monitor the result of such queries over time. We enable users to set a provenance query as: (i)~pull query, where a time-tracker will be assigned to this query. Time-tracker will trigger the start of the querying process at specific user-defined intervals; or (ii)~push query, where a database trigger will be assigned to the entities in the query result. Future changes applied to these entities and their relationships will result in re-executing the query. The result of pull and push queries can be stored in timed folder and path nodes. An intelligent agent\footnote{Intelligent Agent (IA) is an autonomous entity which observes and acts upon an environment and directs its activity towards achieving goals.} will be created and allocated to a timed folder/path node in order to monitor its evolution and update its content. Users can initialize an intelligent agent in order to set a time interval or to assign it to a database trigger.

\section{Example Scenario}
\label{bgScenario}

\begin{figure} [t]
\centering
  \includegraphics[scale=0.5]{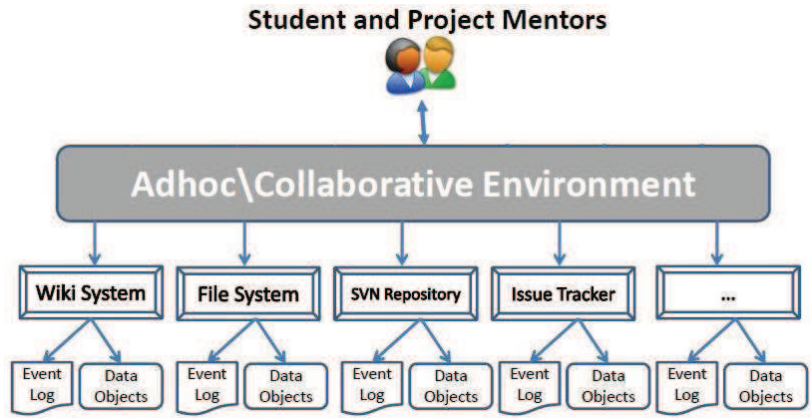}\\
  \caption{Motivating Scenario.}\label{MotivatingScenario}
\end{figure}

To understand the limitations of current approaches in modeling provenance metadata, with respect to time, we present an example scenario. The provenance metadata, in this scenario, collected according to the Open Provenance Model (OPM)~\cite{Moreau:OPM, OPM2010}. OPM is designed to be a standard graph data model for provenance. Currently, OPM does not support temporal properties among the first class abstractions, i.e. OPM graph nodes and edges, but time information can be added as optional annotations to the first class objects. This makes analyzing and querying provenance information represented in OPM, with respect to time, challenging and cumbersome. To show this, we compare our approach with that of querying OPM models where time is considered as annotation. Moreover, in section~\ref{convAlgo}, we present an algorithm to generate a TPM graph from an existing OPM graph. This algorithm can be used to support systems already adapted to OPM.

We choose a motivating scenario from the business intelligence domain, and build on our experience on managing an online project-based course ``e-Enterprise Projects"~\footnote{http://www.cse.unsw.edu.au/$\sim$cs9323} during 2009 and 2010. There were people from different groups (e.g. students, mentors and lecturers) involved in this course. For example in the 2nd semester 2009 we had 66 people (60 students + 5 project mentors + 1 lecturer) involved in course activities. During this semester, fifteen project groups (each group consists of four students) were formed where each group was allocated to one of the available projects. Each mentor was allocated to supervise three projects. The development process of each project went through a sequence of pre-defined  phases:  brainstorming, requirements analysis, design phase, prototype implementation, testing, and final product delivery.

\begin{figure}
  \centering
  \includegraphics[scale=0.6]{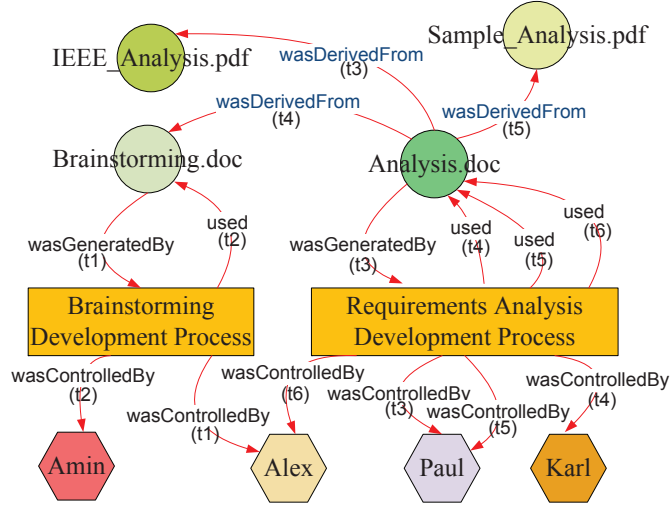}\\
  \caption{OPM representation for the Example~1.}\label{fig:OPM}
\end{figure}

\begin{figure}
  \centering
  \includegraphics[scale=0.6]{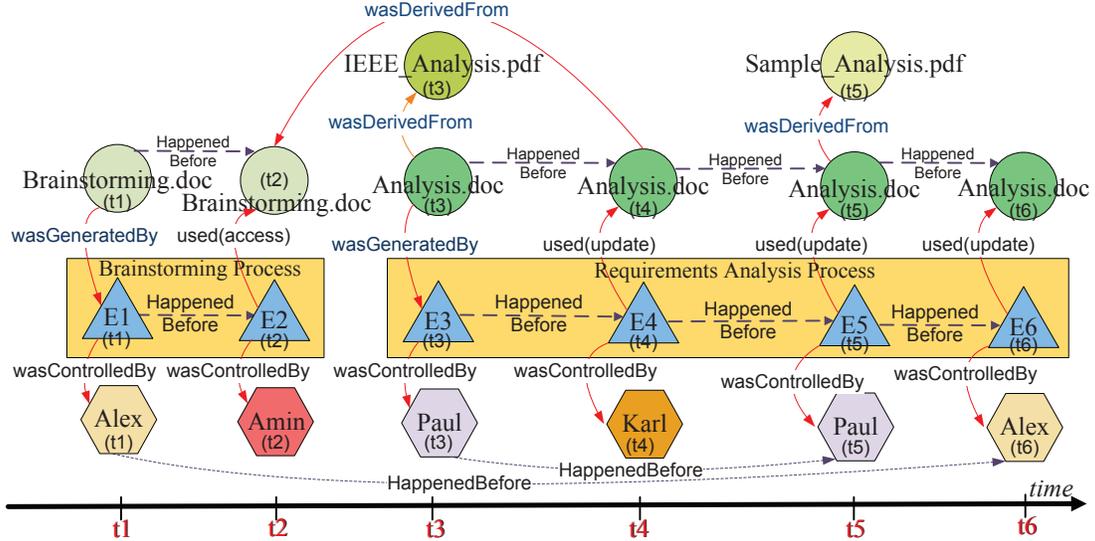}\\
  \caption{TPM representation for the Example~1.}\label{fig:TPM}
\end{figure}

The activities of each project was documented according to open provenance model (OPM) and through a web-based project management system which was equipped with many back-end modules such as: 1) Message board: to exchange message and open discussion topics between the project members. 2) Wiki system: which is used to collaboratively edit related documents to the activities of their project. 3) Blogging system: where each user has their own blog to edit their own posts. 4) File sharing system: where project members can share access to different files and documents. 5) SVN repository: to synchronize the editing of the projects source codes.  Figure~\ref{MotivatingScenario} depicts an illustration of our motivating scenario. In this scenario, we should note that none of the back-end module can be fully-aware of the entire executed processes. Each back-end module observes only a partial view of the whole executed processes. Moreover, the back-end modules can not force any control on the activities of the running projects.
\\\\
\textbf{Example~1.} Consider a project, e.g. project number four, in e-Enterprise course whose members are Alex, Paul, and Karl and mentor is Amin. For the brainstorming phase of the project, all the members and the mentor attend the brainstorming group meeting. At time $t_1$, Alex was nominated to document the discussion and generate a brainstorming document (i.e. a file named `Brainstorming.doc') on the project management Website. As an ongoing process, at time $t_2$, Amin (the mentor) accesses the Website to assess the brainstorming document. As next series of actions, at time $t_3$, Paul accesses the Website and uses the IEEE-analysis documentation file (which was uploaded by the mentor on the Website) to generate the template draft of the requirements analysis document (i.e. a file named `Analysis.doc'). At time $t_4$, Karl accesses the Website and uses `Brainstorming.doc' file to update the `Analysis.doc' file. During analysis development process, students asked the mentor to provide a sample requirements-analysis document (i.e. a file named `Sample$\_$Analysis.pdf'). At time $t_5$, Paul accesses the Website and use `Sample$\_$Analysis.pdf' file to update `Analysis.doc' file. At time $t_6$, Alex accesses the Website to update the `Analysis.doc' file (e.g. for proof reading and polishing). Figures~\ref{fig:OPM} and~\ref{fig:TPM}  illustrate the OPM graph and TPM graph representation for provenance metadata in this example, respectively. We will use and explain these two graphs throughout the rest of the paper.

\section{TPM: T\textbf{emporal} P\textbf{rovenance} M\textbf{odel}}
\label{SecMetaModel}

We assume that provenance of data objects is represented by a directed acyclic graph. We define Temporal Provenance Model (TPM) as a graph data model for organizing a set of entities as graph nodes and entity relationships as edges of the provenance graph in time. An entity is a data object that exists independently and has a unique identity. TPM supports: (i)~structured and unstructured entities; (ii)~folder nodes, which contain entity collections. A folder node represent the results of a query that returns a collection of related entities; and (iii)~path nodes, which refer to one or more paths in the graph, which are the result of a query. A path is the result of the transitive relationship between two entities. Timed folder/path nodes can collect and maintain the complete history or ancestry of their evolution over time. In TPM, entities and relationships are represented as a directed graph $G_{(\tau_1,\tau_2)} =(V_{(\tau_1,\tau_2)}, E_{(\tau_1,\tau_2)})$, where $V_{(\tau_1,\tau_2)}$ is a set of nodes representing instances of entities in time or timed folder/path nodes added to the graph between a time period of $\tau_1$ and $\tau_2$, and $E_{(\tau_1,\tau_2)}$ is a set of directed edges representing relationships among nodes. It is possible to capture the evolution of the graph $G_{(\tau_1,\tau_2)}$ between a time period of $\tau_1$ and $\tau_2$. Detailed data model definitions of path and folder nodes can be found in~\cite{FPSPARQLtechrep,BPM11}.

\subsection{TPM Entities}
\label{Entities}

Entities can be structured or unstructured. Structured entities are instances of entity types. An entity type consists of a set of attributes. Unstructured entities, are also described by a set of attributes but may not conform to an entity type. This entity model offers flexibility when types are unknown and takes advantage of structure when types are known. For the sake of simplicity, we assume all unstructured entities are instances of a generic type called \emph{ITEM}. ITEM is similar to \emph{generic table} in \cite{OneTableStoresAll}. In TPM, entities are represented by a set of instances, each for a given point in time. For example, entity $E$ is represented by the set of instances $\{E_{t_1},E_{t_2},E_{t_3},...\}$ where $\{t_1,t_2,t_3,...\}$ indicates the activity timestamps at distinct points in time. We assume that each distinct activity does not have a temporal duration. Entity instances considered as data objects that exist separately and have a unique identity (e.g. entity unique identity and an activity timestamp). Different instances of an entity, for different points in time, may have different attribute values. For example a digital file may have different content at two different points in time.
\\\\
\textbf{Definition 1. [Event]} is a record of an activity in a system. We consider events as the finest, most granular and basic source of information. Events may be related to one another~\cite{Luckham:CEP}. In principle, every IT system usually generates events that describe the execution of its activities. When the activity starts, the event is created and a reading from a (centralized) clock is entered into the event as its timestamp. The order of timestamps defines the temporal dependency among them. Events can belong to different event types (e.g. repository update or editing blog post in section~\ref{bgScenario}). Each recorded event is usually documented with its $EventID$ and creation timestamp in addition to a set of descriptive attributes. These attributes are different from one event type to another. For example, in the motivating scenario, a message board event is usually documented by metadata attributes which represents the sender of the message, the receiver of the message, the message subject, and the message content while a repository update event can be described by who is making the update in addition to descriptive comments for the update.
\\\\
\textbf{Definition 2. [Artifact]} defined as product of human intelligence and effort, and is classified as tangible (e.g. a physical object) and intangible (e.g. digital representation of an object). In principle, the execution of any event can have an attached list of artifacts. For example, in the motivating scenario, $Alice$ sends $John$ a reminder message about the project meeting with an attached PDF file that describes the agenda of the meeting. $Alice$ can also create an issue to solve a programming bug with a set of screenshots that describes the sequence of actions (i.e. process) which generate the bug. In TPM, artifact $A$ is represented by the set of instances $\{A_{t_1},A_{t_2},A_{t_3},...\}$ where $\{t_1,t_2,t_3,...\}$ indicate the activity timestamps at distinct points in time. For example, in Figure~\ref{fig:TPM}, artifact `analysis.doc' is represented by the set of instances $\{'analysis.doc'_{t_3},'analysis.doc'_{t_4},'analysis.doc'_{t_5},'analysis.doc'_{t_6}\}$ where $\{t_3,t_4,t_5,t_6\}$ indicate the activity timestamps applied on this artifact over time.
\\\\
\textbf{Definition 3. [Agent]} is needed to perform the execution of any work item, e.g. an activity (i.e. event) or series of related activities (i.e. process). An agent is an entity which is capable of acting as a catalyst of an event or process~\cite{Luckham:CEP}. Agents can be either human or non-human (e.g. software application, web service, hardware device). Each agent is usually described by a set of metadata attributes. For example, a human resource can be described by its role, qualification, skills and any other personal attributes. In TPM, agent $Ag$ is represented by the set of instances $\{Ag_{t_1},Ag_{t_2},Ag_{t_3},...\}$ where $\{t_1,t_2,t_3,...\}$ indicate the activity timestamps, e.g. the activity that required the agent for a given point in time to complete its execution. For example, in Figure~\ref{fig:TPM}, agent `Alex' is represented by the set of instances $\{Alex_{t_1},Alex_{t_6}\}$ where $\{t_1,t_6\}$ indicates the activity timestamps at distinct points~in~time.
\\\\
\textbf{Definition 4. [Timed Folder Node]} defined as a timed container for a set of related entities which are connected through \emph{transitive} or \emph{non-transitive} relationships. In other words, the set of entities in a folder node is the result of a given change-aware query that requires grouping graph entities in a certain way. The change-aware query, documents the evolution of folder node by adapting an intelligent agent (see section~\ref{GraphQueryLanguages}). A folder creates a higher level node on top of which other queries could be executed. Folders can be nested, i.e. be members of another folder node, to allow creating and querying folders with relationships at higher levels of abstraction. A folder may have a set of attributes that describe it. A folder node is added to the graph and can be stored to enable reuse of the query results for frequent or recurrent queries. Entities and relationships in a timed folder node are represented as a subgraph $F_{(\tau_1,\tau_2)} =(V_{(\tau_1,\tau_2)}, E_{(\tau_1,\tau_2)})$, where $V_{(\tau_1,\tau_2)}$ is a set of related nodes representing instances of entities in time added to the folder $F$ between a time period of $\tau_1$ and $\tau_2$, and $E_{(\tau_1,\tau_2)}$ is a set of directed edges representing \emph{transitive} and/or \emph{non-transitive} relationships among these related nodes. It is possible to capture the evolution of the folder $F_{(\tau_1,\tau_2)}$ between a time period of $\tau_1$ and $\tau_2$.

Folder nodes can belong to different types, e.g. type process (i.e. a folder can represent a process instance). Each folder node is documented with its ID, starting timestamp, duration (i.e. the length of time it represents) and a set of descriptive attributes. Timed folder node $F$ can be represented by the set of instances $\{F_{(t_1,d_1)},F_{(t_2,d_2)},F_{(t_3,d_3)},...\}$ where $\{t_1,t_2,t_3,...\}$ indicate the starting activity timestamps and $\{d_1,d_2,d_3,$ $...\}$ represents the temporal duration for each folder. For example, a \emph{process} (i.e. series of events performed on or caused by artifacts, and may result in new artifacts) is a composed data object that exists separately, has a unique identity, and can be represented as a timed folder node. A process can be structured or semi-structured. Often, a formal (i.e. structured) or an informal (i.e. semi-structured) description of the process is available in the form of a process graph which can be modeled as a folder (see previous works of authors~\cite{BPM11,hamid2010} as an example). Processes can belong to different process types, e.g.~brainstorming process process. Process $PS$ can be represented by set of instances $\{PS_{(t_1,d_1)},PS_{(t_2,d_2)},PS_{(t_3,d_3)},...\}$ where $\{t_1,t_2,t_3,...\}$ indicate the process starting timestamps, and $\{d_1,d_2,d_3,...\}$ indicate process instances duration.
\\\\
\textbf{Definition 5. [Timed Path Node]} defined as a timed container for a set of related entities which are connected through \emph{transitive} relationships (i.e. a path is a transitive relationship between two entities showing the sequence of edges from the start entity to the end). This relationship can be codified using regular expressions~\cite{BPM11} in which alphabets are the nodes and edges from the graph. We define a timed path node for each change-aware query which results in a set of paths. The change-aware query, documents the evolution of path node by adapting an intelligent agent (see section~\ref{GraphQueryLanguages}). We use existing reachability approaches to verify whether an entity is reachable from another entity in the graph. Some reachability approaches (e.g. all-pairs shortest path~\cite{CodebookPath}) report all possible paths between two entities. Entities and relationships in a timed path node are represented as a subgraph $P_{(\tau_1,\tau_2)} =(V_{(\tau_1,\tau_2)}, E_{(\tau_1,\tau_2)})$, where $V_{(\tau_1,\tau_2)}$ is a set of related nodes representing instances of entities in time which added to the path node $P$ between a time period of $\tau_1$ and $\tau_2$, and $E_{(\tau_1,\tau_2)}$ is a set of directed edges representing \emph{transitive} relationships among these related nodes. It is possible to capture the evolution of the path node $P_{(\tau_1,\tau_2)}$ between a time period of $\tau_1$ and $\tau_2$.

Path nodes can belong to different types, e.g. \emph{derivation} which is the path(s) starting from an instance of an artifact in time to its origin(s). Each path in a timed path node as well as the timed path node itself, is documented with its ID and a set of descriptive attributes. Timed path node $P$ can be represented by the set of instances $\{P_{(t_1,d_1)},$ $P_{(t_2,d_2)},$ $P_{(t_3,d_3)},...\}$ where $\{t_1,t_2,t_3,...\}$ indicate the starting activity timestamps and $\{d_1,d_2,d_3,...\}$ represents the temporal duration for each path Node. For example, consider an analyst who is interested in analyzing the provenance graphs for different versions of an artifact. Different versions of artifact $A$ can be stored in path nodes and represented by the set of instances $\{A_{{v_1}_{(t_1,d_1)}},$ $A_{{v_2}_{(t_2,d_2)}},$ $A_{{v_3}_{(t_3,d_3)}},$ $...\}$ where $v$ represents the version number, $t$ represents the starting activity timestamp of each version, and $d$ represents the temporal duration for each version life-cycle. The provenance graph for each version of an artifact can be stored and used for further analysis.

\begin{figure} [t]
  \centering
  \includegraphics[scale=0.64]{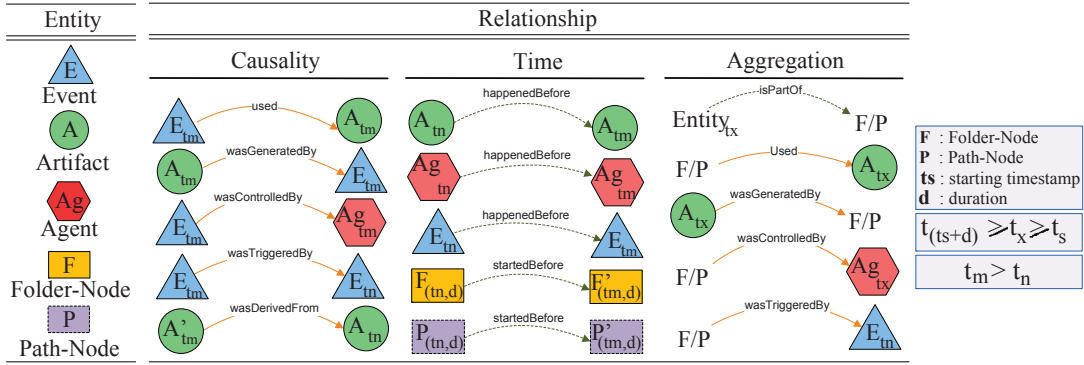}\\
  \caption{Temporal Provenance Model (TPM) notations.}\label{FigHypergraphRepresentation}
\end{figure}

We introduce a graphical notation and a formal definition for TPM graphs (Figure \ref{FigHypergraphRepresentation}). Specifically, events are represented by triangles, artifacts are represented by circles, agents are represented by octagons, folder nodes are represented by rectangles, and path nodes are represented by dashed rectangles.

\subsection{TPM Relationships}
\label{Relationship}

A relationship is a directed link between a pair of entities, which is associated with a predicate defined on the attributes of entities that characterizes the relationship. A relationship can be \emph{explicit} (e.g. `Event-6'$\rightarrow$`used'$\rightarrow$`analysis.doc' in Figure~\ref{fig:TPM}) or \emph{implicit}, such as a relationship between an entity and a larger (composite) entity. In TPM, we introduce three types of relationships among graph nodes: \emph{Causality}, \emph{Time}, and \emph{Aggregation}.

\subsubsection{\large Causality}

Causality is a dependency relationship among activities in a system. An action (i.e. event) or series of actions (i.e. process) can perform on or cause artifacts, and result in new artifacts. Today's approaches in modeling provenance try to model the causality dependency among system objects. For example, five causal relationships are recognized in OPM~\cite{OPM2010,Moreau:OPM}: an event `used' an artifact, an artifact `wasGeneratedBy' an event, an event `wasTriggeredBy' an event, an artifact `wasDerivedFrom' an artifact, and an event `wasControlledBy' an agent. We leveraged these relationships to capture the causal dependencies among TPM entities.
\\\\
\textbf{Definition 6.} [Event $E_{t_m}$ \emph{used} Artifact $A_{t_m}$] In a graph, connecting an event to an artifact by a `used' edge is intended to indicate that: (a) the event required the availability of the artifact to complete its execution; (b) $t_m$ is the event $E$ timestamp; and (c) $A_{t_m}$ is an instance of artifact $A$ at time $t_m$ used by event $E$.
\\\\
\textbf{Definition 7.} [Artifact $A_{t_m}$ \emph{wasGeneratedBy} Event $E_{t_m}$] In a graph, connecting an artifact to an event by an edge `wasGeneratedBy' is intended to indicate that: (a) the event was required to initiate its execution for the artifact to be generated; (b) $t_m$ is the event $E$ timestamp; and (c) $A_{t_m}$ is the first instance of artifact $A$ generated at time $t_m$.
\\\\
\textbf{Definition 8.} [Event $E_{t_m}$ \emph{wasTriggeredBy} Event $E_{t_n}$] A connection of an event $E_{t_m}$ to an event $E_{t_n}$ by a `wasTriggeredBy' edge indicates that: (a) the start of event $E_{t_n}$ was required for $E_{t_m}$ to be able to complete; and (b) $t_m > t_n$.
\\\\
\textbf{Definition 9.} [Artifact $A'_{t_m}$ \emph{wasDerivedFrom} Artifact $A_{t_n}$] The assertion of an edge `wasDerivedFrom' between two artifacts $A'_{t_m}$ and $A_{t_n}$ indicates that: (a) artifact $A'_{t_m}$ was generated by event $E_{t_m}$; (b) event $E_{t_m}$ used artifact $A_{t_n}$ to generate artifact $A'_{t_m}$; and (c) $t_m > t_n$.
\\\\
\textbf{Definition 10.} [Event $E_{t_m}$ \emph{wasControlledBy} Agent $Ag_{t_m}$] The assertion of an edge `wasControlledBy' between an Event $E_{t_m}$ and an agent $Ag_{t_m}$ indicates that the execution of event $E_{t_m}$ was controlled by agent $Ag$ at time ${t_m}$.
\\\\
\textbf{Definition 11.} A folder/path node ($F_{(t_s,d)}$/$P_{(t_s,d)}$) can inherit all incoming and outgoing causal dependencies among its member entities $\{En_1,$ $En_2,$ $En_3,$ $...,$ $En_i\}$ and other entities in the TPM graph. Four causal relationships can be inherited: (i) a folder/path node $F_{(t_s,d)}$/$P_{(t_s,d)}$  `used' an artifact $A_{t_m}$, where $(t_s+d) \geq t_m \geq t_s$; (ii) an artifact $A_{t_m}$ `wasGeneratedBy' a folder/path node $F_{(t_s,d)}$/$P_{(t_s,d)}$, where $(t_s+d) \geq t_m \geq t_s$; (iii) a folder/path node $F_{(t_s,d)}$/$P_{(t_s,d)}$ `wasTriggeredBy' an event $E_m$, where $t_m < t_s$; and (iv) a folder/path node $F_{(t_s,d)}$/$P_{(t_s,d)}$ `wasControlledBy' an agent $A_{t_m}$, where $(t_s+d) \geq t_m \geq t_s$.

For example, if process $P_{(t_s,d)}$ consists of set of events $\{E_1,E_2,E_3,...,E_i\}$, then $P_{(t_s,d)}$ can inherit all incoming and outgoing causal dependencies of its members $\{E_1,E_2,E_3,...,$ $E_i\}$, which are: (i)~a process $P_{(t_s,d)}$ `used' an artifact $A_{t_m}$, where $(t_s+d) \geq t_m \geq t_s$; (ii)~an artifact $A_{t_m}$ `wasGeneratedBy' a process $P_{(t_s,d)}$, where $(t_s+d) \geq t_m \geq t_s$; (iii)~a process $P_{(t_s,d)}$ `wasTriggeredBy' an event $E_m$, where $t_m < t_s$; and (iv)~a process $P_{(t_s,d)}$ `wasControlledBy' an agent $A_{t_m}$, where $(t_s+d) \geq t_m \geq t_s$. These relationships have the similar definition as explained in event causality dependencies and emphasize that a process, during its life-cycle, may: (a)~use series of artifacts; (b)~generate series of artifacts; and (c)~be controlled by series of agents (see figure~\ref{fig:TPM}).

\subsubsection{\large Time}

Time is the relationship that is used to order activities and instances of entities in a chronological manner based on their occurrence. Kostakos~\cite{Kostakos:TemporalGraph} proposes to express the temporal dependency between two entities using the `happened before' relationship which is a directed edge of weight `$t_m$-$t_n$' where $t_m$ and $t_n$ represent timestamps of activities or instances of entities in time, and $t_m>t_n$. We leveraged this relationship to connect related events and instances of artifacts and agents over periods of time. Moreover, to connect related folder/path nodes, we use `started before' relationship which is a directed edges of weight `$t_m$-$t_n$' where $t_m$ and $t_n$ represent the starting activity timestamps of entities in a folder/path node,~and~$t_m>t_n$.
\\\\
\textbf{Definition 12.} [Event $E_{t_m}$ \emph{happenedBefore} Event $E_{t_n}$] In a graph, connecting an event to another event by a `happened before' edge is intended to indicate that: (a)~the first event ($E_{t_m}$) happened before the second event ($E_{t_n}$); and (b)~these events have a specific dependency (e.g. they are part of series of related activities, i.e., a process); and (c) \{$E_{t_m}$,$E_{t_n}$\} is a consecutive pair and `$t_n$-$t_m$' represents the temporal distance between the elements of the pair.
\\\\
\textbf{Definition 13.} [Artifact $A_{t_m}$ \emph{happenedBefore} Artifact $A_{t_n}$] In a graph, connecting an artifact to another artifact by a `happened before' edge is intended to indicate that: (a)~artifact $A$ is represented by the set of instances {$A_{t_1}$, $A_{t_2}$, ..., $A_{t_n}$} where $n$ represents the number of instances over time; (b)~$A_{t_m}$ is an instance of artifact $A$ at time $t_m$; (c)~$A_{t_n}$ is an instance of artifact $A$ at time $t_n$; and (d)~\{$A_{t_m}$,$A_{t_n}$\} is a consecutive pair and `$t_n$-$t_m$' represents the temporal distance between the elements of the pair.
\\\\
\textbf{Definition 14.} [Agent $Ag_{t_m}$ \emph{happenedBefore} Agent $Ag_{t_n}$] In a graph, connecting an agent to another agent by a `happened before' edge is intended to indicate that: (a)~agent Ag is represented by the set of instances (i.e. states) {$Ag_{t_1}$, $Ag_{t_2}$, ..., $Ag_{t_n}$} where $n$ represents the number of instances over time; (b)~$Ag_{t_m}$ is an instance of agent Ag at time $t_m$; (c)~$Ag_{t_n}$ is an instance of agent Ag at time $t_n$; and (d)~\{$Ag_{t_m}$,$Ag_{t_n}$\} is a consecutive pair and $t_n$-$t_m$ represents the temporal distance between the elements of the pair.
\\\\
\textbf{Definition 15.} [Folder-Node $F_{(t_m,d)}$ \emph{startedBefore} Folder-Node $F_{(t_n,d)}$] In a graph, connecting a folder node to another folder node by a `started before' edge is intended to indicate that: (a)~Folder-Node $F$ is represented by the set of instances {$F_{(t_1,d_1)}$, $F_{(t_2,d_2)}$, ..., $F_{(t_n,d_n)}$} where $n$ represents the number of instances of folder $F$ over periods of time, $d$ (duration) represents the temporal distance between the staring and ending activity timestamp, and $t$ represent the starting activity timestamp in the instance; (b)~$F_{(t_m,d)}$ is an instance of folder node $F$ at time $t_m$; (c)~$F_{(t_n,d)}$ is an instance of folder node $F$ at time $t_n$; (d)~the starting timestamp of $F_{(t_m,d)}$ is smaller than the starting timestamp of $F_{(t_n,d)}$, sorted in chronological order; and (e)~\{$F_{(t_m,d)}$,$P_{(t_n,d)}$\} is a consecutive pair and $t_n$-$t_m$ represents the starting activity timestamp of the pair. For example, different instances of a process (i.e. represented as folders) can be connected through a `started before' relationship.
\\\\
\textbf{Definition 16.} [Path-Node $P_{(t_m,d)}$ \emph{startedBefore} Path-Node $P_{(t_n,d)}$] In a graph, connecting a path node to another path node by a `started before' edge is intended to indicate that: (a)~Path-Node $P$ is represented by the set of instances {$P_{(t_1,d_1)}$, $P_{(t_2,d_2)}$, ..., $P_{(t_n,d_n)}$} where $n$ represents the number of instances of path node $P$ over periods of time, $d$ represents the temporal distance between the starting and ending activity timestamp for each instance, and $t$ represent the starting activity timestamp in the instance; (b)~$P_{(t_m,d)}$ is an instance of path node $P$ at time $t_m$; (c)~$P_{(t_n,d)}$ is an instance of path node $P$ at time $t_n$; (d)~the starting timestamp of $P_{(t_m,d)}$ is smaller than the starting timestamp of $P_{(t_n,d)}$, sorted in chronological order; and (e)~\{$P_{(t_m,d)}$,$P_{(t_n,d)}$\} is a consecutive pair and $t_n$-$t_m$ represents the temporal distance between the starting activity timestamp of the pair. For example, the path(s) from different snapshots (versions) of an artifact to their origin(s) can be stored in different path nodes where two consecutive path node can be connected through a `started before' relationship.

\subsubsection{\large Aggregation}

Aggregation is an abstraction relationship. It signifies a complex entity that the aggregated entities signify~\cite{Luckham:CEP}. Aggregated entities are hierarchically organized by `is part of' (i.e. an implicit relationship) relationships, and can be stored in a folder/path node. Processes are examples of such complex entities. If process $P_{(t_s,d)}$ signifies an activity that consists of the activity of a set of events $\{E_1,E_2,E_3,...,E_i\}$, then $P_{(t_s,d)}$ is an aggregation of all the events $E_i$.
\\\\
\textbf{Definition 17.} [Entity $En_{t_m}$ \emph{isPartOf} FolderNode $F_{(t_s,d)}$] In a graph, connecting a TPM entity $En_{t_m}$ (i.e. event, artifact, agent, folder node, and path node) to a folder node $F_{(t_s,d)}$ by an edge `is part of' is intended to indicate that: (a) entity $En_{t_m}$ is part of aggregated entities which signify folder node $F_{(t_s,d)}$; and (b) $(t_s+d) \geq t_m \geq t_s$.
\\\\
\textbf{Definition 18.} [Entity $En_{t_m}$ \emph{isPartOf} PathNode $P_{(t_s,d)}$] In a graph, connecting a TPM entity $En_{t_m}$ (i.e. event, artifact, agent, folder node, and path node) to a path node $F_{(t_s,d)}$ by an edge `is part of' is intended to indicate that: (a) entity $En_{t_m}$ is part of aggregated entities which signify path node $P_{(t_s,d)}$; and (b) $(t_s+d) \geq t_m \geq t_s$.

\subsection{OPM-to-TPM Conversion Algorithm}
\label{convAlgo}

\textbf{Open Provenance Model.} Prior work on modeling and representing provenance takes various forms under the names of lineage, pedigree, or tracing~\cite{provSurvey1,provSurvey2,provSurvey3}. OPM (Open Provenance Model~\cite{Moreau:OPM}) is designed to be a standard graph data model for provenance. Emerging models for provenance from a wide range of domains map well to terms and extensibility mechanisms defined in OPM. Moreover, W3C provenance team~\cite{websiteOPMW3C} chose to adopt OPM V1.1 as the provenance target model since it is already a community model, which has undergone several revisions, and which is adopted by different systems (e.g. VisTrails\footnote{http://www.vistrails.org/index.php/Main$\_$Page} and Karma\footnote{http://pti.iu.edu/d2i/provenance$\_$karma}). In OPM, digital representation of provenance is in the form of an annotated causality graph, i.e. a directed acyclic graph, which is based on three primary entities: artifact, process, and agent. OPM aims to capture the causal dependencies between abovementioned entities (i.e. graph nodes) by introducing five types of causality dependencies: a process `used' an artifact, artifacts `generated by' processes, process `triggered by' process, artifact `derived from' artifact, and process `controlled by' agent. OPM does not specify protocol for storing or querying provenance information. Moreover, annotations are not part of the vocabulary (i.e. artifacts, processes, agents, and the casual dependencies between them) provided by OPM. Even though the open provenance model is timeless~\cite{Moreau:OPM}, it allows for causality graphs to be annotated with time annotations. Kwasnikowska et.~al.~\cite{OPM2010} proposed a formal definition for temporal semantics for OPM graphs, i.e. the annotated causality graph, defined in terms of a set of ordering constraints between time-points associated with OPM constructs. These constraints allow OPM graphs to be decorated with time information needed for the beginning of a process, the ending of a process, the instant a process uses an artifact, and the moment a process creates an artifact. In such annotated graph, analyzing and querying relationships between nodes over time and understanding the role of each node in the temporal context of the entire graph becomes complex and cumbersome, at best.
\\\\
\textbf{Conversion Algorithm.} To support systems already using OPM, we propose an algorithm to generate a TPM graph from an existing OPM graph. We translate annotated temporal information to generate instances of entities over time. For example, consider artifact $A_2$ in Figure~\ref{OPM_TPM_conv}(a). According to the fact that this artifact was used at times $t_2$ and $t_4$, the conversion algorithm will generate two instances for artifact $A_2$ per point in time, $A_{2_{t2}}$ and $A_{2_{t4}}$ (see Figure~\ref{OPM_TPM_conv}(b)), in TPM graph. As another example consider process $P2$ in Figure~\ref{OPM_TPM_conv}(a). According to the fact that this process generated artifact $A_1$ (at time $t_1$) and used artifact $A_2$ (at time $t_2$), the conversion algorithm will generate two events, $E_{t1}$ and $E_{t2}$ (see Figure~\ref{OPM_TPM_conv}(b)), as members of this process in TPM~graph.

An algorithm to convert a graph into its equivalent temporal graph has been proposed in \cite{Kostakos:TemporalGraph}. This algorithm does not support the conversion of complex entities, i.e. set of related entities (e.g.~process instances). We extend this algorithm to discover related entities in OPM graph, group them in folder/path nodes of type process, and consider them as graph nodes in generating the TPM graph. For example, consider the OPM graph representation of Example~1 (Figure~\ref{fig:OPM} in section~\ref{bgScenario}). Figure~\ref{fig:TPM} in section~\ref{bgScenario} illustrates the TPM graph representation for this example generated by applying the following OPM-to-TPM graph conversion algorithm in three steps:

\begin{figure}
  \centering
  \includegraphics[scale=0.32]{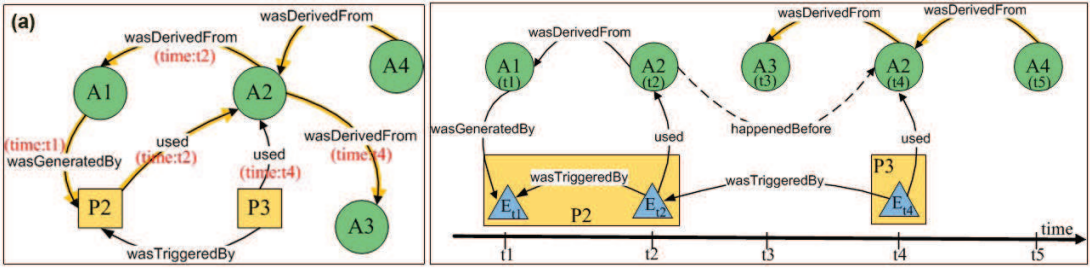}\\
  \caption{An OPM graph (a) and its equivalent TPM graph (b) in the motivating~scenario.}\label{OPM_TPM_conv}
\end{figure}

\begin{enumerate}
  \item Identifying the role of each \emph{artifact} in the temporal context of the graph:
    \begin{itemize}
      \item Create one node per artifact per point in time (i.e. activity timestamp). So Artifact $A$ is represented by the set of instances \{$A_{t_1}$, $A_{t_2}$, ..., $A_{t_n}$\} where: (a) $n$ represents the number of instances of artifact $A$ over periods of time; and (b) $\{t_1,t_2,t_3,...\}$ indicates the activity timestamps at distinct points in time.
      \item For each set of artifact instances we link consecutive pairs \{$A_{t_x}$, $A_{t_{x+1}}$\} with `happened before' edges of weight $t_{x+1} - t_x$ (definition~13), representing the temporal distance between the pair.\\
    \end{itemize}

  \item Identifying the role of each \emph{agent} in the temporal context of the graph:
    \begin{itemize}
      \item Create one node per agent per point in time (i.e. activity timestamp). So Agent $Ag$ is represented by the set of instances \{$Ag_{t_1}$, $Ag_{t_2}$, ..., $Ag_{t_m}$\} where: (a) $m$ represents the number of instances of agent $Ag$ over periods of time; and (b) $\{t_1,t_2,t_3,...\}$ indicates the activity timestamps at distinct points in time.
      \item For each set of agent instances we link consecutive pairs \{$Ag_{t_x}$, $Ag_{t_{x+1}}$\} with `happened before' edges of weight $t_{x+1} - t_x$ (definition~14), representing the temporal distance between the pair.\\
    \end{itemize}

  \item Identifying the role of \emph{activities} in the temporal context of the graph:
    \begin{itemize}
      \item Discover process instances and group them in folder/path nodes. This can be done through existing work (the previous work of authors~\cite{BPM11,hamid2010}).
      \item For each set of related events in a process instances we link consecutive pairs \{$E_{t_x}$, $E_{t_{x+1}}$\} with `happened before' edges of weight $t_{x+1} - t_x$ (definition~12), representing the temporal distance between the pair.
      \item For each set of process instances we link consecutive pairs \{$P_{(t_x,d)}$, $P_{(t_{x+1},d)}$\} with `started before' edges of weight $t_{x+1} - t_x$ (definition~15-16), representing the temporal distance between the pair.
      \item Use unweighted directed edges (definition 6-11) to link artifact and agent instances that participated in an activity process. The result will be various instances of artifacts and agents per point in time, reconnected to activity processes.
      \item Use aggregation relationships (definition~17-18) to link folder/path nodes to their members.
    \end{itemize}

\end{enumerate}

\section{Querying TPM Graphs}
\label{SecQueryLanguage}

Querying TPM graphs needs a graph query language that not only supports primitive graph queries but also is capable of: (i)~constructing timed folder/path nodes. In general, the output of every query can be stored as timed folder/path nodes and used for further querying. For example, a query can be used to discover path(s) to the origin of the artifact `analysis.doc' in Example~1. The output of this query can be stored as a timed path node and used for further querying; (ii)~applying further queries to constructed timed folder/path nodes. For example, consider a timed path node which contains the lifecycle of the artifact `analysis.doc' in Example~1. We can apply further queries to this timed path node to analyze: (a)~the evolution of this node, or (b)~the activities performed on or caused by this artifact at different points and periods of time; and (iii)~applying external tools and algorithms to TPM graphs for further analysis. For example, we may need to apply existing reachability algorithms, graph mining tools/algorithm, or frequent pattern discovery algorithm to TPM graphs.

We proposed FPSPARQL~\cite{BPM11} (a Folder-Path enabled extension of SPARQL~\cite{SPARQL}), a graph query processing engine, which supports primitive graph queries, constructing folder/path nodes, applying further queries to constructed folder/path nodes, and applying external tools and algorithms to graph. There are two levels of queries in FPSPARQL: (a)~Graph-level Queries: at this level SPARQL is used to query graphs; and (b)~Node-level Queries: at this level FPSPARQL extends SPARQL to construct and query folder/path nodes. In particular, $fconstruct$ statement in FPSARQL groups a set of related entities or folders and stores the result, i.e. a subgraph, under a folder node name. The $pconstruct$ statement in FPSARQL finds transitive relationships, i.e. paths, between entities and store them under a path node name. Finally, FPSARQL's $apply$ statement retrieves information, i.e. by applying queries, from the underlying folder and path nodes. FPSPARQL does not support the construction and querying of \emph{timed} folder/path nodes. Following we extend FPSPARQL query engine to address this requirement.

\subsection{Constructing Timed Folder Nodes}
\label{CQTimedFolderNodes}

\begin{figure*}
\centering
  \includegraphics[scale=0.6]{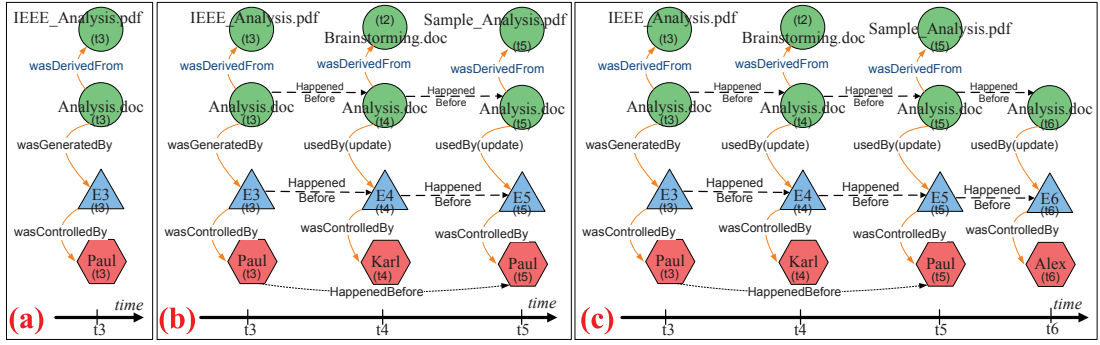}\\
  \caption{The content of the times folder node constructed in Example~2 at time: (a)~$t_3$, (b)~$t_5$, and (c)~$t_6$.}\label{folder_evolution}
\end{figure*}

To construct a timed folder node, we extend FPSPARQL's $fconstruct$ statement. This command is used to group a set of related entities or folders. The syntax for a basic construction query of a timed folder node is given as follows:

\begin{verbatim}
fconstruct <Folder_Node Name> as ?folder
[select ?var1 ?var2 ... | (Folder1, Folder2,...)]
where {
 ?folder @timed true.
 (other patterns)
}
\end{verbatim}

A query can be used to define a new timed folder node by listing folder node name and entity definitions in the \emph{fconstruct} and \emph{select} statements, respectively. Also a folder node can be defined to group a set of folder nodes. A set of user-defined attributes for this folder can be defined in the \emph{where} statement. Setting the value of attribute $timed$ to $true$ for the folder, will assign an intelligent agent to this folder. The intelligent agent is responsible for updating the folder content over periods of time.
\\\\
\textbf{Example~2.} Considering e-Enterprise course scenario, a timed folder can be constructed in order to monitor `all the activities having happened in the requirements analysis phase of group number four' (see Example~1). We assume that requirement analysis activities will take place between timestamps $t_3$ and $t_6$. New activities would be automatically updated in this folder as time passes by. Figure~\ref{folder_evolution} shows the content of this folder at times $t_3$, $t_5$, and $t_6$. Following is a sample FPSPARQL query for this example.

\begin{verbatim}
fconstruct analysis_process as ?anlPrs
select ?e
where {
 ?anlPrs @timed true.
 ?anlPrs @isA folderNode.
 ?anlPrs @type process.
 ?anlPrs @description `analysis activities'.
 ?e @isA entityNode.
 ?e @type event. ?e @timestamp
 ?ts. FILTER ( Timesemantic(?ts, [t3,?,?,t6]) ).
}
\end{verbatim}

In this example, variable $?anlPrs$ represents the folder node to be constructed, i.e. `analysis\_process'. The first four patterns in $where$ statement define folder attributes (e.g. timed, type, and description). Variable $?e$ represents the activities to be stored in the folder, and variable $?ts$ represents the timestamp of these activities.

\begin{table} [b]
  \centering
  \caption{FPSPARQL time semantics.}\label{tbl:timeSemantics}
  \scalebox{0.9}{
  \begin{tabular}{cc}
  \\
  \hline
  Time Semantic & Time Range \\ [0.3ex]
  \hline
    in, on, at, during          & [t,t,t,t] \\
    since                       & [t,t,?,?] \\
    after                       & [t,?,?,?] \\
    before                      & [?,?,?,t] \\
    till, untill, by            & [?,?,t,t] \\
    between                     & [t,?,?,t] \\
  \hline
  \end{tabular}
  }
\end{table}

For the sake of simplicity in writing temporal queries, we adapted the time semantics proposed in~\cite{TOB}. We introduce the special construct,
$timesemantic(fact,[t1,t2,t3,t4])$ in FPSPARQL, which is used to represent the $fact$ to be in a specific time interval $[t1,t2,t3,t4]$. A fact may have no temporal duration (e.g. a distinct event) or may have temporal duration (e.g. series of activities stored in a folder node and considered as a process instance). Table~\ref{tbl:timeSemantics} represents the time-semantics that we support in FPSPARQL queries. In this example, $timesemantic$ statement defines events timestamps, i.e. variable $?ts$, to be between timestamps $t3$ and $t6$, i.e. start and end of requirement analysis phase.

\subsection{Querying Timed Folder Nodes}

Using the $apply$ statement in FPSPARQL, it is possible to apply queries to constructed timed folder nodes. For example, consider that we want to apply the query `give me the list of artifacts that ``$Analysis.doc$" file was derived from during process $P_{(t,d)}$ activities' to the folder constructed in Example~2. Following is a sample FPSPARQL query for this example.

\begin{verbatim}
(analysis_process) apply (
 select ?docID
 where {
  ?a @isA entityNode.
  ?a @type artifact.
  ?a @id `Analysis.doc'.
  ?a @timestamp ?ts.
  ?a wasDerivedFrom ?a2.
  ?a2 @id ?docID.
  filter(Timesemantic(?ts,[t,?,?,t+d])).
 }
)
\end{verbatim}

The $apply$ command, in FPSPARQL, is used to retrieve information, i.e. by applying queries, from the underlying folder/path nodes. In this example the query applied to the folder `analysis\_process'. Variable $?a$ represents the artifact `$Analysis.doc$' and variable $?a2$ represents the artifacts that `$Analysis.doc$' was derived from. The $timesemantic$ statement defines the timestamps of instances of `$Analysis.doc$', i.e. variable $?ts$, to be between the starting time (i.e. $t$) and ending time (i.e. $t+d$) of process $P_{(t,d)}$.
\\\\
\textbf{Example~3. [Querying the Evolution of Timed Folder Nodes]} Using the $apply$ statement in FPSPARQL, it is possible to retrieve information about folder evolution in time. Consider a user who is interested to observe the information of the constructed folder in Example~2 at time $t_5$. Following is the FPSPARQL query for this example.

\begin{verbatim}
(analysis_process) apply (
 select *
 where {
  ?a @isA entityNode.
  ?a @timestamp ?ts.
  filter( Timesemantic(?ts,[?,?,?,t5]) ).
 }
)
\end{verbatim}

In this example the query applied to the constructed timed folder node `analysis\_process'. Variable $?a$ represents all members of the folder node whose timestamps $?ts$ fall before time $t_5$. Figure~\ref{folder_evolution}(b) illustrates the result of this query. Similar queries can be used to retrieve the content of this folder between two specific timestamps $\tau_1$ and $\tau_2$, e.g. by replacing filter statement with $filter( Timesemantic(?ts,[\tau_1,?,?,\tau_2]) )$ in the above example.

\subsection{Constructing Timed Path Nodes}
\label{CQTimedPathNodes}

\begin{figure*}
  \centering
  \includegraphics[width=1.0\textwidth]{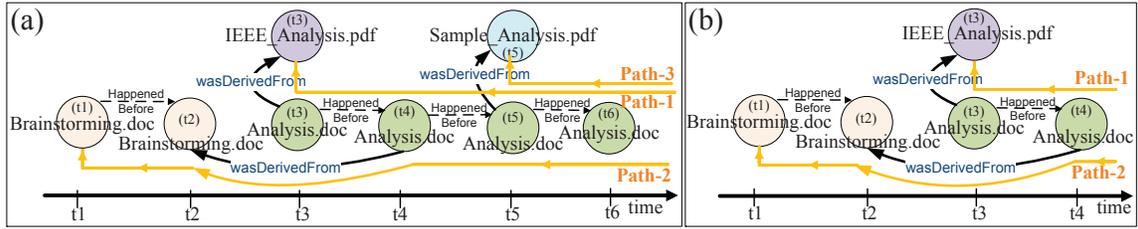}\\
  \caption{Examples of constructing and querying timed path nodes: (a) the result of derivation query in Example~5, and (b) the result of querying the evolution of timed path node in Example~6} \label{ArtifactDerivation}
\end{figure*}

To construct a timed path node, we extend the $pconstruct$ command in FPSPARQL. This command is used to discover transitive relationships between two entities and store it under a path node name. The syntax for a basic construction query of a timed path node is given as follows:

\begin{verbatim}
pconstruct <Path_Node Name>
(StartNode,EndNode,RegularExpression) as ?pathNode
where {
 ?pathNode @timed true.
 (other patterns)
}
\end{verbatim}

A regular expression can be used to define a transitive relationship between two entities~\cite{BPM11}. Attributes of starting node, ending node, and regular expression's alphabets (i.e. graph nodes and edges) can be defined in the \emph{where} statement. Setting the value of attribute $timed$ to $true$, will assign an intelligent agent to this path node. In~\cite{BPM11}, we showed that $pconstruct$ queries can be executed without specifying `start node' and `end node' parameters in order to discover frequent patterns. We introduce two examples of timed path nodes which are useful in analyzing provenance graphs: \emph{time-series} and \emph{derivation}.
\\\\
\textbf{Example~4. [Time-series]} We define a \emph{time-series} as a transitive relationship among instances of a TPM entity over periods of time, showing the sequence of `happened before' edges. Considering e-Enterprise course scenario, it is possible to construct a timed path node in order to monitor `the time-series of ``Analysis.doc" file in Example~1'. Following is a sample FPSPARQL query for this example.

\begin{verbatim}
pconstruct analysisDoc_timeseries
( , ,?node (?edge ?node)+) as ?anlDocTS
where {
 ?anlDocTS @timed true.
 ?anlDocTS @isA pathNode.
 ?anlDocTS @type timeseries.
 ?anlDocTS @description `artifact timeseries'.
 ?node @isA entityNode.
 ?node @Type artifact.
 ?node @id `Analysis.doc'.
 ?edge @isA edge.
 ?edge @label happenedBefore.
}
\end{verbatim}

In this example, the $pconstruct$ statement returns the pattern to be discovered and stores it under a path node name. Variable $?anlDocTS$ represents the path node to be constructed, i.e. `analysisDoc\_timeseries'. The first four patterns in $where$ statement define path attributes (e.g. timed, type, and description). Setting the value of attribute $timed$ to $true$, will assign an intelligent agent to this path node. In the regular expression $?node (?edge$ $?node)+$, parentheses are used to define the scope and precedence of the operators, the plus sign indicates there is one or more of the preceding element, and the asterisk indicates there are zero or more of the preceding element. Variable $?node$ indicates the artifact `$Analysis.doc$' and variable $?edge$ indicates directed edges labeled `happened before'.
\\\\
\textbf{Example~5. [Derivation]} We define a \emph{derivation} as the \emph{ancestry relationships}\cite{provenanceQuery1} starting from instances of an artifact to its origin(s), showing the sequence of `happened before' and `was derived from' edges. According to this definition, derivation of an artifact is a set of time-lines to its origin. Considering e-Enterprise course scenario, a timed path node can be constructed in order to monitor `the derivation of ``Analysis.doc" file in Example~1'. Following is a sample FPSPARQL query for this example.

\begin{verbatim}
pconstruct analysisDoc_derivation
( , ,?artifact (?edge ?artifact)+ (?e ?n)*) as ?anlDocDRV
where {
 ?anlDocDRV @isA pathNode.
 ?anlDocDRV @timed true.
 ?anlDocDRV @type derivation.
 ?anlDocDRV @description `artifact derivation'.
 ?artifact @isA entityNode.
 ?artifact @id `Analysis.doc'.
 ?edge @isA edge.
 ?edge @label happenedBefore.
 ?n @isA entityNode.
 ?e @isA edge.
 ?e @label ?label.
 FILTER (?label=wasDerivedFrom || ?label=happenedBefore).
}
\end{verbatim}

In this example, variable $?anlDocDRV$ represents the path node to be constructed, i.e. `analysisDoc\_derivation'. The first four patterns in $where$ statement define path attributes. Setting the value of attribute $timed$ to $true$, will assign an intelligent agent to this path node. In the regular expression $?artifact (?edge$ $?artifact)+$ $(?e$ $?n)*$, variable $?artifact$ indicates the artifact `Analysis.doc' and variable $?edge$ indicates directed edges labeled `happened before'. Variable $?n$ can indicate any entity node and variable $?e$ indicates directed edges labeled `was derived from' or `happened before'. Figure \ref{ArtifactDerivation}(a) illustrates the result of this query which includes three paths to the origins of `Analysis.doc'.

\subsection{Querying Timed Path Nodes}

Using the $apply$ statement in FPSPARQL, it will be possible to apply queries to the constructed path nodes. For example, we can apply the query `Retrieve the time-series of ``Analysis.doc" file, in the time period between $t_3$ and $t_5$' to the path node constructed in Example~4. Following is a sample FPSPARQL query for this example.

\begin{verbatim}
(analysisDoc_timeseries) apply (
 select * where {
  ?a @isA entityNode.
  ?a @timestamp ?ts.
  filter( Timesemantic(?ts,[t3,?,?,t5]) ).
 })
\end{verbatim}

In this example the query applied to the constructed timed path node in Example~4, i.e. `analysisDoc\_timeseries'. Variable $?a$ represents all members of the path node whose timestamps $?ts$ fall between timestampes $t_3$ and $t_5$.
\\\\
\textbf{Example~6. [Querying the Evolution of Timed Path Nodes]} Using the $apply$ statement in FPSPARQL, it is possible to retrieve information about path nodes evolution at different points in time. Consider a user who is interested to see the information of the path node constructed in Example~5 (i.e. derivation) at time $t_4$. Following is a sample FPSPARQL query for this example.

\begin{verbatim}
(analysisDoc_derivation) apply (
 select *
 where {
  ?a @isA entityNode.
  ?a @timestamp ?ts.
  filter( Timesemantic(?ts,[?,?,?,t4]) ).
 }
)
\end{verbatim}

In this example the query applied to the constructed path node in Example~5. This path node consists of three paths (see Figure~\ref{folder_evolution}(a)). The query will apply to each of the three paths, and the result will be generated for each path individually. Figure~\ref{folder_evolution}(b) illustrates the result of this query. In this example, variable $?a$ represents all members of the path node whose timestamps $?ts$ falls before time $t_4$.

\section{Architecture, Implementation and Experiments}
\label{Experiments}

\subsection{Architecture}

\begin{figure} [t]
  \centering
  \includegraphics[width=0.8\textwidth]{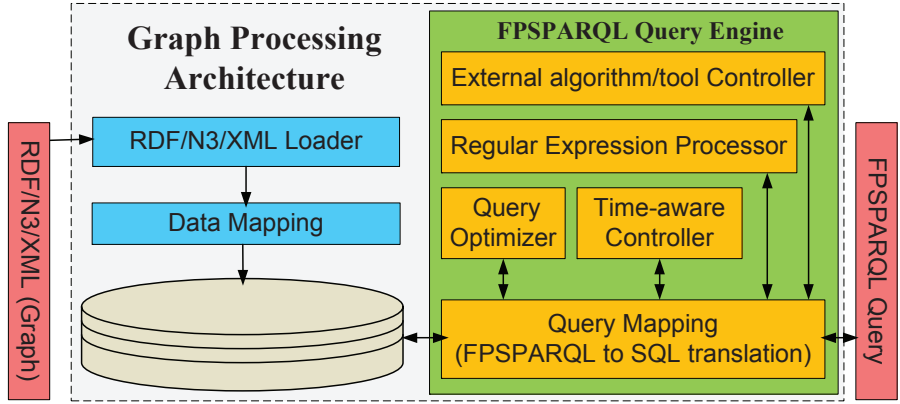}\\
  \caption{FPSPARQL graph processing architecture.} \label{graphArchitecture}
\end{figure}

Figure~\ref{graphArchitecture} illustrates FPSPARQL graph processing architecture which consists of following components:

\begin{enumerate}
  \item \emph{Graph Loader}: Input graph can be in the form of RDF, N3 (or Notation3, is a W3C standard and shorthand non-XML serialization of RDF models), or XML. We developed a workload-independent physical design by developing a \emph{loader} algorithm. This algorithm is responsible for: (i)~validating the input graph; (ii)~generating the relational representation of triple store, for manipulating and querying entities, folders, and paths; and (iii)~generating powerful indexing mechanisms.

  \item \emph{Data Mapping Layer}: is responsible for creating data element mappings between semantic web technology (i.e. Resource Description Framework) and relational database schema.

  \item \emph{Query Mapping Layer}: is consist of a FPSPARQL parser (for parsing FPSPARQL queries based upon the syntax of FPSPARQL) and a schema-independent FPSPARQL-to-SQL translation algorithm. This algorithm consists of:

      \begin{itemize}
        \item ~\emph{SPARQL-to-SQL Translation Algorithm.} We implemented a SPARQL-to-SQL translation algorithm based on the proposed relational algebra for SPARQL~\cite{SPARQLAlgebra} and semantics preserving SPARQL-to-SQL query translation~\cite{SPARQLSemantics}. This algorithm supports \emph{Aggregate} queries and \emph{Keyword Search} queries.
        \item \emph{Folder Node Construction and Querying.} We use the relational representation of triple RDF store, to store, manipulate, and query folder nodes.
        \item \emph{Path Node Construction and Querying.} To describe constraints on the path nodes, we reused the specification for regular expressions and filter expressions proposed in CSPARQL~\cite{RE_path}.
      \end{itemize}

  \item \emph{Regular Expression Processor}: is responsible for parsing the described patterns through the nodes and edges in the graph. We developed a regular expression processor which supports optional elements (?), loops (+,*), alternation (|), and grouping ((...)).

  \item \emph{External Algorithm/Tool Controller}: is responsible for supporting applying external graph reachability algorithm or mining tools to the graph.

  \item \emph{Time-aware Controller}: is responsible for creating an intelligent agent and allocate it to a timed folder/path node in order to monitor its evolution and update its content. We enable users to set a provenance query as: (i)~pull query, where a time-tracker will be assigned to this query. Time-tracker will trigger the start of the querying process at specific user-defined intervals; or (ii)~push query, where a database trigger will be assigned to the entities in the query result. Future changes applied to these entities and their relationships will result in re-executing the query. Users can initialize an intelligent agent in order to allocate it to a timed folder/path node and set its time interval or assign it to a database trigger.

  \item \emph{Query Optimizer}: To optimize the performance of queries, we developed four optimization techniques proposed in \cite{RDFProv,SherifSurvey1,SPARQLSemantics}: (i) selection of queries with specified varying degrees of structure and spanning keyword queries; (ii) selection of the smallest table to query based on the type information of an instance; (iii) elimination of redundancies in basic graph pattern based on the semantics of the patterns and database schema; and (iv) create separate tables (property tables) for subjects that tend to have common properties to reduce the self-join problem.
\end{enumerate}

\subsection{Implementation}

\begin{figure} [t]
\centering
  \includegraphics[scale=0.27]{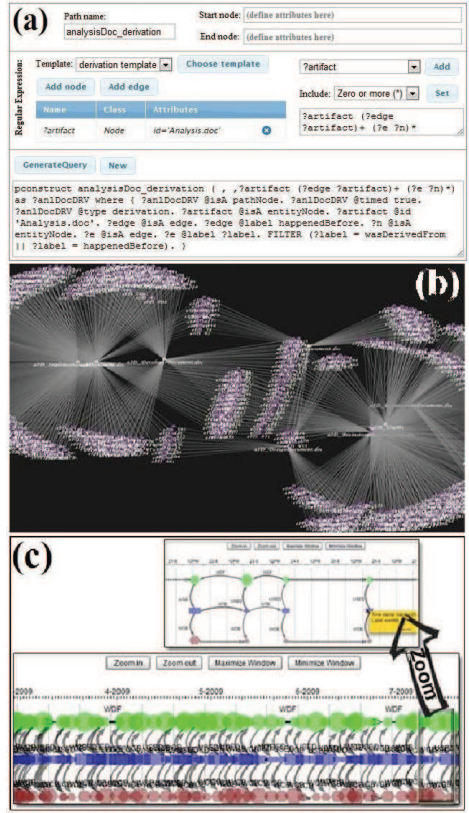}\\
  \caption{Screenshots of front end tool: (a) Query assistant tool: generating the derivation query in Example~5; (b) OPM graph visualization tool: visualization of (part of) e-Enterprise course dataset OPM graph; and (c) TPM graph visualization tool: visualization of (part of) e-Enterprise course dataset TPM graph.
}\label{fig:vis}
\end{figure}

We have implemented a large-scale graph processing engine, i.e. FPSPARQL, and the full details of our data model and query engine are presented in \cite{BPM11,FPSPARQLtechrep}. The simplest way to store a set of RDF statements is to use a relational database with a single table that includes columns for subject, property and object. While simple, this schema quickly hits scalability limitations~\cite{SherifSurvey1}. To avoid this we developed a relational RDF store including its three classification approaches~\cite{SherifSurvey1}: vertical (triple), property (n-ary), and horizontal (binary). The query engine is implemented in Java (J2EE). Moreover, a front-end tool prepared to assist users in two steps:
\\\\
\textbf{Step1:} [Query Assistant] We provide users with a query assistant tool to generate FPSPARQL queries in an easy way. We provide templates for time-series, derivation, and some other useful path construction queries. Moreover, users can use the tool to generate the regular expressions and path queries they are interested in. Figure~\ref{fig:vis}(a) illustrates a screenshot of this tool while generating the derivation query in Example~5.
\\\\
\textbf{Step2:} [Visualizing] We provided users with a graph visualization tool for the exploration of graphs and query results (see Figure~\ref{fig:vis}(b,c)). For the TPM graph exploration, we provide users with a timeline like interface (see Figure ~\ref{fig:vis}(c)) with facilities such as zooming in and out.

\subsection{Datasets}
\label{Datasets}

We carried out the experiments on 3 datasets:

\begin{itemize}
  \item \textbf{e-Enterprise Course.} This dataset was introduced in the motivating scenario in section~\ref{MotivatingScenario}.
  \item \textbf{SCM.} This dataset is the interaction log of a SCM business service, developed based on the supply chain management scenario provided by WS-I (the Web Service Interoperability organization). SCM dataset contains 4,050 events, 14 service operations, and 28 attributes.
  \item \textbf{PurchaseNode.} This process log was produced by a workflow management system supporting a purchase order management service. The log contains 34,803 messages, 26 service operations, and 26 attributes.
\end{itemize}

\begin{table} [b]
\caption{Details for our datasets, and their respective OPM and TPM graphs.}
\centering
\begin{tabular}{ | l || l || l || l || l || l | }

\cline{3-6}
       \multicolumn{2}{ c }{}  &  \multicolumn{2}{|c||}{\cellcolor{black}{\color{white}OPM Graph}} & \multicolumn{2}{|c|}{\cellcolor{black}{\color{white}TPM Graph}} \\
\hline \hline
\multicolumn{1}{|c||}{\cellcolor{black}{\color{white}dataset}}  &
\multicolumn{1}{|c||}{\cellcolor{black}{\color{white}events}}   &
\multicolumn{1}{|c||}{\cellcolor{black}{\color{white}Nodes}}    &
\multicolumn{1}{|c||}{\cellcolor{black}{\color{white}Edges}}    &
\multicolumn{1}{|c||}{\cellcolor{black}{\color{white}Nodes}}    &
\multicolumn{1}{|c|}{\cellcolor{black}{\color{white}Edges}} \\
\hline \hline
\multicolumn{1}{|c||}{\cellcolor{gray}{e-Enterprise Course}} & 104,050 & $\sim 261k$ & $\sim 1,532k$ & $\sim 853k$ & $\sim 2,001k$ \\
\hline \hline
\multicolumn{1}{|c||}{\cellcolor{gray}{SCM}}                 & 4,050   & $\sim 10k$  & $\sim 72k$    & $\sim 48k$  & $\sim 335k$ \\
\hline \hline
\multicolumn{1}{|c||}{\cellcolor{gray}{PurchaseNode}}        & 34,803  & $\sim 83k$  & $\sim 730k$   & $\sim 181k$ & $\sim 1,023k$ \\
\hline \hline
\end{tabular}
\label{tblDatasets}
\end{table}

We applied a preprocessing phase to adapt SCM and PurchaseNode datasets to OPM graph data model. Using this conversion algorithm proposed in section~\ref{convAlgo}, we generate the analogous TPM graph for these datasets. Table \ref{tblDatasets} represents the number of collected events, nodes and edges in OPM graph, and nodes and edges in TPM graph (after conversion) in these datasets. Details about these datasets can be found in the previous work of authors~\cite{hamid2010}.

\subsection{Evaluation}
\label{Evaluation}

\begin{figure} [t]
\centering
  \includegraphics[scale=0.4]{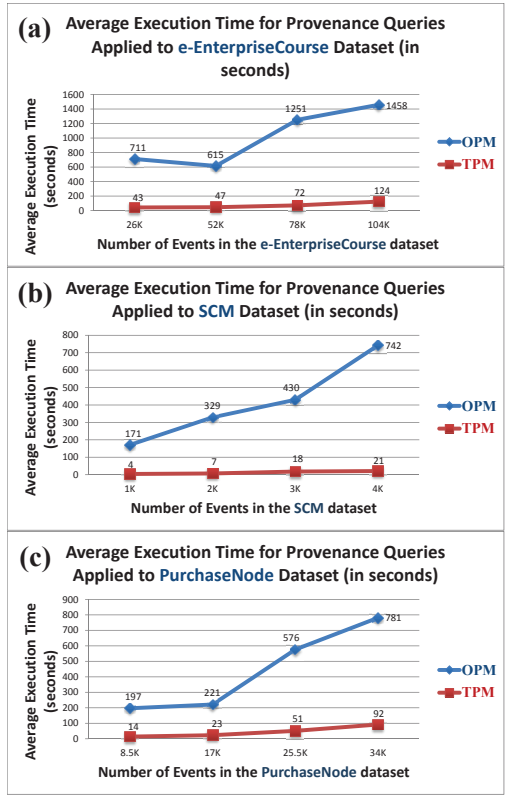}\\
  \caption{The query performance evaluation results, illustrating the average execution time for applying provenance queries on TPM and OPM graph generated from (a)~e-Enterprise course dataset, (b)~SCM dataset, and (c)~PurchaseNode dataset.
}\label{queryPerformance}
\end{figure}

\begin{table}
\caption{Details for different sizes of datasets used in the evaluation including number of events in the datasets and number of TPM/OPM graph nodes generated from them.}
\centering
\begin{tabular}{ | c || c || c || c | }

\hline \hline
\multicolumn{1}{|c||}{\cellcolor{black}{\color{white}}}  &
\multicolumn{1}{|c||}{\cellcolor{black}{\color{white}number of}}   &
\multicolumn{1}{|c||}{\cellcolor{black}{\color{white}number of Nodes}}    &
\multicolumn{1}{|c||}{\cellcolor{black}{\color{white}number of Nodes}}  \\

\multicolumn{1}{|c||}{\cellcolor{black}{\color{white}dataset}}  &
\multicolumn{1}{|c||}{\cellcolor{black}{\color{white}events}}   &
\multicolumn{1}{|c||}{\cellcolor{black}{\color{white}in TPM graph}}    &
\multicolumn{1}{|c||}{\cellcolor{black}{\color{white}in OPM graph}}  \\
\hline \hline
\multicolumn{1}{|c||}{\cellcolor{gray}{ }} & 26K & $\sim 48k$ & $\sim 271k$ \\
\hline
\multicolumn{1}{|c||}{\cellcolor{gray}{e-Enterprise Course}} & 52K & $\sim 133k$ & $\sim 516k$ \\
\hline
\multicolumn{1}{|c||}{\cellcolor{gray}{ }} & 78K & $\sim 153k$ & $\sim 562k$ \\
\hline
\multicolumn{1}{|c||}{\cellcolor{gray}{ }} & 104K & $\sim 261k$ & $\sim 853k$ \\
\hline \hline
\multicolumn{1}{|c||}{\cellcolor{gray}{ }}                 & 1K   & $\sim 3k$  & $\sim 17k$    \\
\hline
\multicolumn{1}{|c||}{\cellcolor{gray}{SCM}}                 & 2K   & $\sim 5k$  & $\sim 21k$    \\
\hline
\multicolumn{1}{|c||}{\cellcolor{gray}{ }}                 & 3K   & $\sim 7k$  & $\sim 42k$    \\
\hline
\multicolumn{1}{|c||}{\cellcolor{gray}{ }}                 & 4K   & $\sim 10k$  & $\sim 48k$    \\
\hline \hline
\multicolumn{1}{|c||}{\cellcolor{gray}{ }}        & 8.5K  & $\sim 18k$  & $\sim 34k$   \\
\hline
\multicolumn{1}{|c||}{\cellcolor{gray}{PurchaseNode}}        & 17K  & $\sim 51k$  & $\sim 117k$   \\
\hline
\multicolumn{1}{|c||}{\cellcolor{gray}{ }}        & 25.5  & $\sim 57k$  & $\sim 160k$   \\
\hline
\multicolumn{1}{|c||}{\cellcolor{gray}{ }}        & 34K  & $\sim 83k$  & $\sim 181k$   \\
\hline \hline
\end{tabular}
\label{tblDS_TPM_OPM}
\end{table}

\begin{table} [t]
\caption{The quality evaluation of query results on three datasets.}
\centering
\begin{tabular}{ | c || c || c || c || c | }

\cline{2-5}
       \multicolumn{1}{ c }{}  &  \multicolumn{2}{|c||}{\cellcolor{black}{\color{white}OPM Graph}} & \multicolumn{2}{|c|}{\cellcolor{black}{\color{white}TPM Graph}} \\
\hline \hline
\multicolumn{1}{|c||}{\cellcolor{black}{\color{white} }}  &
\multicolumn{1}{|c||}{\cellcolor{black}{\color{white} \small Number of}}    &
\multicolumn{1}{|c||}{\cellcolor{black}{\color{white} \small Number of}}    &
\multicolumn{1}{|c||}{\cellcolor{black}{\color{white} \small Number of}}    &
\multicolumn{1}{|c|}{\cellcolor{black}{\color{white} \small Number of}} \\

\multicolumn{1}{|c||}{\cellcolor{black}{\color{white} dataset}}  &
\multicolumn{1}{|c||}{\cellcolor{black}{\color{white} \small discovered paths}}    &
\multicolumn{1}{|c||}{\cellcolor{black}{\color{white} \small relevant paths}}    &
\multicolumn{1}{|c||}{\cellcolor{black}{\color{white} \small discovered paths}}    &
\multicolumn{1}{|c|}{\cellcolor{black}{\color{white} \small relevant paths}} \\

\multicolumn{1}{|c||}{\cellcolor{black}{\color{white} }}  &
\multicolumn{1}{|c||}{\cellcolor{black}{\color{white} \small for all the queries}}    &
\multicolumn{1}{|c||}{\cellcolor{black}{\color{white} }}    &
\multicolumn{1}{|c||}{\cellcolor{black}{\color{white} \small for all the queries}}    &
\multicolumn{1}{|c|}{\cellcolor{black}{\color{white} }} \\
\hline \hline
\multicolumn{1}{|c||}{\cellcolor{gray}{e-Enterprise Course}} & 139 & 41 & 81  & 73  \\
\hline \hline
\multicolumn{1}{|c||}{\cellcolor{gray}{SCM}}                 & 85  & 64 & 68  & 65  \\
\hline \hline
\multicolumn{1}{|c||}{\cellcolor{gray}{PurchaseNode}}        & 94  & 42 & 34  & 33  \\
\hline \hline
\cline{2-5}
       \multicolumn{1}{ c }{ }  &  \multicolumn{2}{|c||}{\cellcolor{gray}{precision=46.2\%}} & \multicolumn{2}{|c||}{\cellcolor{gray}{precision=93.4\%}} \\
\end{tabular}
\label{tblqualityEval}
\end{table}

We have compared our approach with that of querying OPM models. We evaluated the performance and the query results quality using the proposed datasets. Moreover, the performance of FPSPARQL query engine has been evaluated in~\cite{BPM11,FPSPARQLtechrep}. All experiments were conducted on a HP system with a 2.67Ghz Quad CPU, 4 GBytes of memory, and running a 64-bit Win~7.
\\\\
\textbf{Performance.} We evaluated the performance of provenance queries using \emph{execution time} metric. To evaluate the performance of queries, we provided 60 provenance queries (10 queries for OPM graphs and 10 queries for TPM graphs generated for each dataset). These queries were generated by domain experts who were familiar with the proposed datasets and include: (a)~why queries, to specify the influences that a source data had on the existence of the data; (b)~how queries: to specify the action or series of actions performed on or caused by source data; (c)~where queries (i.e. derivation queries): to specify the origin(s) of data; and (d)~when queries: to execute why, how, and where queries over different periods of time. For each query, we generated an equivalent query to be applied to the OPM graphs as well as the TPM graphs for each dataset. As a result, a set of historical paths for each query were discovered and stored in path nodes. Figure~\ref{queryPerformance} shows the average execution time for applying these queries to the OPM graph and the equivalent TPM graph generated from each dataset. As illustrated in Figure~\ref{queryPerformance} we divided each dataset into regular number of events, then generated TPM and OPM graph for different sizes of datasets, and finally ran the experiment for different sizes of TPM and OPM graphs. Table~\ref{tblDS_TPM_OPM} shows the number of events for different sizes of datasets, and the number of graph nodes for each TPM and OPM graph generated from them. The evaluation shows a polynomial (nearly linear) increase in the execution time of the queries with respect to with the dataset size.
\\\\
\textbf{Quality.} The quality of the results is assessed using classical \emph{precision} metric which is defined as the percentage of
discovered results that are actually interesting. For evaluating the interestingness of the result, we asked domain experts who had the most accurate knowledge about the datasets and the related process to analyze discovered paths and identify what they considered relevant and interesting from a provenance perspective. Table~\ref{tblqualityEval} represents the quality evaluation of query results on three datasets. The table illustrates the number of discovered paths for all the provenance queries (generated for performance evaluation) and the number of relevant paths chosen by domain experts. As a result of applying provenance queries to OPM graphs, 318 paths discovered, examined by domain experts, and 147 paths (precision=46\%) were considered relevant. And as a result of applying provenance queries to TPM graphs, 183 paths were discovered and examined by domain experts, and 171 paths (precision=93.4\%) were considered relevant.
\\\\
\textbf{Discussion.} We evaluated our approach using different time-sensitive datasets and compared TPM with that of querying OPM models. As evaluation shows, for each query applied to a TPM graph and its equivalent OPM graph, the number of discovered paths in an OPM graph is much more than the number of discovered paths in the equivalent TPM graph, i.e. path queries applied to OPM graphs resulted in many irrelevant paths. We discovered many cycles in the results of path queries applied to OPM graphs (and to eliminate these cycles, we applied the cycle elimination techniques proposed in~\cite{GraphBook}). Provenance metadata can be collected at different levels of granularity and shifting granularity risks the creation of cyclic provenance~\cite{provCycle}, e.g. provenance can be collected for an artifact, its multiple snapshots over time, or groups of related artifacts. Figure~\ref{queryPerformance} shows that performing time-aware queries based on an OPM model will lead to a decreased query performance and an increased complexity to analyze and understand temporal provenance graph of entity instances. Furthermore, the evaluation shows that, time, intervals, versioning, and node merging will reduce cycles and irrelevant paths in path query results applied to temporal provenance metadata.

We implemented an interface to support various graph reachability algorithms~\cite{GraphBook} such as all-pairs shortest path, transitive closure, GRIPP, tree cover, chain cover, path-tree cover, and Sketch. In general, there are two types of graph reachability algorithms~\cite{GraphBook}: (1) algorithms traversing from starting vertex to ending vertex using breadth-first or depth-first search over the graph, and (2) algorithms checking whether the connection between two nodes exists in the edge transitive closure of the graph. Considering  $G=(V,E)$ as directed graph that has $n$ nodes and $m$ edges, the first approach imposes a time complexity of $O(n+m)$ and the second approach imposes a space complexity of $O(n^2)$. In this experiment, we used the GRIPP (from first category) and all-pairs shortest-paths (from second category) algorithms. In both cases, path queries applied to OPM graphs maximized the consumption of memory and processor and resulted in many irrelevant paths and cycles in the query~result. Overall, evaluation shows the proposed model is an appropriate approach for discovering paths (i.e. basis of many provenance queries) through temporal provenance graphs.

\section{Conclusion}
\label{Conclusion}

In this paper, we have presented a temporal provenance model (TPM) for modeling, querying, and analyzing data provenance. Two concepts of timed folder and path node have been introduced, which help in analyzing temporal provenance graphs. Folders enable grouping related entities and paths help in analyzing the history of entities in time. Timed folder and path nodes show their evolution for the time period they represent. We have extended our previous work, FPSPARQL~\cite{FPSPARQLtechrep,BPM11}, which is a scalable graph query processing engine, to query and analyze TPM~graphs. To evaluate the viability and efficiency of the proposed framework, we have compared our approach with that of querying OPM models where time is considered as annotation. We have conducted experiments over realworld datasets and the evaluation shows the viability and efficiency of our approach. A front-end tool has been provided to facilitate the exploration and visualization of TPM graphs and assisting users with generating provenance queries. As future work, we plan to design a visual query interface to support users in expressing their queries over the conceptual representation of the TPM graph in an easy way. Moreover, we plan to employ interactive graph exploration and visualization techniques (e.g. storytelling systems~\cite{Tolkien}) to help users quickly identify the interesting parts of the graph.

\bibliographystyle{plain}
\bibliography{sample}

\end{document}